\documentstyle[aps,12pt,epsfig]{revtex}
\newcommand{\bm}{\bibitem}

\newcommand{\cp}{\chi^{(+)}}
\newcommand{\cm}{\chi^{(-)*}}

\newcommand{\vv}{V_{bc}({\bf r}_1)}
\newcommand{\ri}{{\bf r}_i}
\newcommand{\ro}{{\bf r}_1}
\newcommand{\ak}{{\bf k}_a}
\newcommand{\bq}{{\bf k}_b}

\newcommand{\rc}{{\bf r}_c}
\newcommand{\cq}{{\bf k}_c}

\begin{document}

\draft
\tighten

\title{Coulomb-nuclear interference in the breakup of 
$^{11}$Be }
\date{\today}
\author{R. Chatterjee and R. Shyam}
\address { 
Saha Institute of Nuclear Physics, 1/AF Bidhan Nagar, Calcutta 700064,
India} 
\maketitle

\begin{abstract}
Within a theory of breakup reactions formulated in the framework
of the post form distorted wave Born approximation, we calculate  
contributions of the pure Coulomb and the pure nuclear breakup as well
as those of their interference terms to a variety of cross sections
in breakup reactions of the one-neutron halo nucleus $^{11}$Be on 
a number of target nuclei. In contrast to the assumption often made,
the Coulomb-nuclear interference terms are found
to be non-negligible in case of exclusive cross sections      
of the fragments emitted in this reaction on medium mass
and heavy target nuclei. The consideration of the nuclear breakup 
leads to a better description of such data.  
\end{abstract}
\pacs{25.60.-t, 25.60.Gc, 24.10.Eq}
 
\newpage

Projectile breakup reactions have played a major role in probing the
structure of neutron rich light radioactive nuclei~\cite{tan88}.
Features of the breakup data such as strongly forward peaked angular
distributions and extremely narrow parallel momentum distributions of
the fragments~\cite{tan88,han95,nak94}, have contributed in a major
way in confirming the existence of a novel structure, called neutron
halo~\cite{han87}, in some of these nuclei. The data on breakup studies
of radioactive nuclei have been increasing rapidly both in quality
and quantity~\cite{tan88,han95,nak94,dav98}. In majority of them both
Coulomb and nuclear breakup effects as well as their interference terms
are likely to be significant. However, in many analyses of the experimental
data on halo breakup reactions the latter term has not been included
\cite{ann94,pra93,mad01,mds01}.        

Therefore, a theoretical treatment of breakup reactions of 
radioactive nuclei, where Coulomb, nuclear, and their
interference terms are treated consistently on an equal footing, is 
an important requirement in efforts to extract the information about
the structure of light exotic nuclei from the experimental data.
For breakup reactions of light stable isotopes, such a theory has
been formulated within the post form distorted wave Born approximation
(DWBA)~\cite{bau84} where this reaction is 
treated as a direct process in which the incoming projectile breaks up
instantaneously in the nuclear and Coulomb fields of the target.
Even though this theory has been remarkably successful in describing
the light ion breakup data~\cite{bau84}, its application to
calculations of breakup of heavier projectiles and at higher
beam energies is not reliable as it uses the simplifying approximation
of a zero-range (ZR) interaction (see, e.g.,~\cite{sat64}) between 
constituents of the projectile. The ZR approximation is inapplicable
in cases where the internal orbital angular momentum of the projectile is
different from zero. Recently, an extended version of this theory where 
the ZR approximation is avoided, has been used to investigate the pure
Coulomb breakup of one- and two-neutron halo nuclei\cite{cha00,cha01}. 

Pure Coulomb and pure nuclear breakup of halo nuclei have been studied
in several different approaches~\cite{nak94,hen96,bon88,yab92}.
On the other hand, in Ref.~\cite{shy99}, the pure nuclear and the
pure Coulomb breakup as well as their interference terms have been
treated on the same footing, in a study of $^8$B breakup within 
a reaction model which describes breakup as an excitation of the
projectile to a two-body continuum state. The corresponding $T$ matrix
is written in terms of the prior form DWBA where interactions between
the fragments and the target are treated in first order.
With this approximation, the prior form DWBA is no longer equivalent to
its post form counterpart~\cite{bau84}. The prior DWBA can be regarded as
the first iteration of the solutions of a coupled channels problem 
(e.g., the coupled discretized continuum channels or CDCC equations).  
In breakup studies of both the stable isotopes~\cite{kam86} and 
halo nuclei~\cite{tos01,mor02}, it is shown that the prior DWBA is
insufficient to describe the data; higher-order coupling effects
of the breakup channels are found to be important in both the cases.
For example, the prior DWBA results for $^8$B breakup at low
beam energies, as shown in~\cite{shy99}, are changed completely by the
higher order effects~\cite{tos01}. For the higher beam energy
($\approx$ 50 MeV/nucleon) case studied in~\cite{shy99},
it is expected \cite{ian02} that coupled channels effects
would be noticeable for angles beyond 5$^\circ$ while in the region
below this they may be relatively weaker.
 
Contributions of the Coulomb and nuclear breakup as well as 
those of their interference terms have also been calculated 
within models~\cite{typ01,bon02} where the time evolution of the
projectile in coordinate space, is described by solving the time
dependent Schr\"odinger equation, treating the projectile-target
(both Coulomb and nuclear) 
interaction as a time dependent external perturbation. These calculations 
use the semiclassical concept of the motion of the projectile along
a trajectory. While in~\cite{typ01} no perturbative
approximation has been made in calculations of the breakup 
cross section, the Coulomb breakup amplitudes have been calculated
in the first order perturbation theory in~\cite{bon02}.   
 
In this paper, we present calculations for breakup of the 
one-neutron halo nucleus $^{11}$Be within the post form DWBA theory
of the breakup reactions that includes consistently both
Coulomb and nuclear interactions between the projectile fragments
and the target nucleus to all orders but treats the fragment-fragment
interaction in first order. This is an extension of the theory
presented in~\cite{cha00} which was able to describe only the pure
Coulomb breakup of such nuclei. Like in~\cite{cha00}, finite
range effects are included within the local momentum
approximation (LMA)~\cite{bra74}. Full ground state wave function of the
projectile of any orbital angular momentum structure, enters into this
theory. It can treat the Coulomb and nuclear breakup as well as their
interference terms consistently within a single setup. Since this theory
uses the post form scattering amplitude, the breakup contributions from the
entire continuum corresponding to all the multipoles and the relative orbital
angular momenta between the valence nucleon and the core fragment are
included into it. Furthermore, 
it can account for the postacceleration effects in an unique
way~\cite{pra02}. Within this theory, we investigate here  
the role of the nuclear and the Coulomb-nuclear interference (CNI)
terms in breakup reactions of the halo nucleus $^{11}$Be.  

We consider the elastic breakup reaction, $a + t \to b + c + t$,
in which the projectile $a$ ($ a = b + c) $ breaks up into 
fragments $b$ and $c$ in the Coulomb and nuclear fields of a target $t$.
Unlike the assumption made in~\cite{cha00}, both fragments can be
charged. The triple differential cross section for this reaction is given by
\begin{eqnarray}
{{d^3\sigma}\over{dE_bd\Omega_bd\Omega_c}} & = &
{2\pi\over{\hbar v_a}}\rho(E_b,\Omega_b,\Omega_c)
\sum_{\ell m}|\beta_{\ell m}|^2,
\end{eqnarray}
where $\rho(E_b,\Omega_b,\Omega_c)$ is the appropriate three-body phase
space factor (e.g., see,~\cite{cha00}), $v_a$ the velocity of $a$, and $\ell$
the orbital angular momentum for the relative motion of $b$ and $c$
in the ground state of $a$. The amplitude $\beta_{\ell m}$ is
defined as 
\begin{eqnarray}
\hat{\ell}\beta_{\ell m}(\bq,\cq;\ak) & = & 
\int d\ro d\ri\cm_b(\bq,{\bf r})\cm_c(\cq,\rc) \vv \nonumber \\
& & \times u_\ell (r_1) Y_{\ell m} ({\hat r}_1)\cp_a(\ak,\ri),
\end{eqnarray}
with ${\hat \ell} \equiv \sqrt{2\ell + 1}$.
In Eq.~(2), functions $\chi_i$ represent 
the distorted waves for the relative motions of various particles 
in their respective channels with appropriate 
boundary conditions. Arguments of these functions contain the
corresponding Jacobi momenta and coordinates.  $\vv$ represents the
interaction between $b$ and $c$, and $u_\ell (r_1)$ the radial part of
the corresponding wave function in the ground state of $a$. 
The position vectors satisfy the relations: 
${\bf r}= \ri - \alpha\ro, {\bf r}_c = \gamma \ro +\delta \ri$,
with $~~ \alpha = (m_c/m_a),\, ~~ \delta = [m_t/(m_b+m_t)]$,
and $~~  \gamma = (1 - \alpha\delta)$. It may be noted that Eq.~(1) 
uses full three-body kinematics and it can readily be used to analyze
the coincidence breakup data (see, e.g.,~\cite {gui00}) of halo nuclei
which are now becoming available with the advent of the secondary beams of
sufficiently high intensity. 

To facilitate an easier computation of Eq.~(2) which involves a
six dimensional integral with the integrand having a product of three
scattering waves that exhibit an oscillatory behavior asymptotically, 
we perform a Taylor series expansion of the distorted waves
of particles $b$ and $c$ about ${\bf r}_i$ and write 
\begin{eqnarray}
\chi^{(-)}_b(\bq,{\bf r}) & = & e^{-i\alpha{\bf K}_b.\ro}
                           \chi^{(-)}_b(\bq,\ri),\\
\chi^{(-)}_c(\cq,{\bf r}_c) & = & e^{i\gamma{\bf K}_c.\ro}
                           \chi^{(-)}_c(\cq,\delta\ri).
\end{eqnarray}
We now employ the LMA~\cite{shy85,bra74}, the attractive feature of
which is that it leads to the factorization of Eq.~(2) into two terms,
each involving a three-dimensional integral. In the LMA, the magnitudes
of momenta ${\bf K}_j$ are taken as $K_j(R)  = \sqrt {(2m_j/ \hbar^2)[E_j
- V_j(R)]}$, where $m_j$ is the reduced mass of the $j-t$ system,
$E_j$ is the energy of particle $j$ relative to the target in the
c.m.\ system, and $V_j(R)$ is the potential between $j$ and 
$t$ at a distance $R$.  As is shown in Ref.~\cite{cha00}, the magnitude
of $K(R)$ remains constant for R $>$ 10 fm for the reaction under
investigation in this paper. Due to the peripheral nature of breakup
reactions, this region contributes maximally to the cross section.
Therefore, we have taken a constant magnitude for $K_j$ evaluated at 
$R$ = 10 fm for all the values of the associated radial variable. 
Furthermore, we checked that the results
of the calculations are almost independent of the choice of the 
direction of the local momentum. Hence, we  take the direction of
${\bf K}_j$ to be the same as that of the asymptotic momentum ${\bf k}_j$.
A detailed discussion of the validity of the LMA, as applied to 
the reaction under investigation here, can be found in~\cite{cha00,zad02}. 

Substituting Eqs.~(3)-(4) into Eq.~(2), introducing the partial wave
expansion of the distorted waves and carrying out the angular momentum 
algebra, we get
\begin{eqnarray}
{\hat l} \beta_{lm} &=&
{(4\pi)^{3} \over {k_a k_b k_c\delta }} i^{-l} Y_{l m}({\hat {\bf Q}})
Z_\ell (Q) \sum_{L_aL_bL_c} (i)^{L_a-L_b-L_c} {\hat L}_b{\hat L}_c
\nonumber \\
& \times & {\cal Y}^{L_b}_{L_c}({\hat k}_b,{\hat k}_c)
 \langle L_b 0 L_c 0| L_a 0 \rangle {\cal R}_{L_b,L_c,L_a}(k_a,k_b,k_a), 
\end{eqnarray}
where
\begin{eqnarray}
{\cal Y}^{L_b}_{L_c}({\hat k}_b,{\hat k}_c) & = &
\sum_M (-)^M\langle L_b M L_c -M|L_a 0 \rangle
Y_{L_b M}({\hat {k}}_b)Y_{L_c M}^*({\hat {k}}_c),\nonumber \\  
Z_\ell (Q) & = & \int_0^{\infty} r_1^2 dr_1 j_{l}(Qr_1) u_l(r_1)V_{bc}(r_1),
\nonumber \\ 
{\cal R}_{L_b,L_c,L_a} & = & \int_0^{\infty}
{{dr_i}\over {r_i}} f_{L_a}(k_a,r_i)f_{L_b}(k_b,r_i) f_{L_c}(k_c,\delta r_i).
\nonumber 
\end{eqnarray}
In Eq.~(5), $Q$ is the magnitude of vector  
${\bf Q} = \gamma {\bf K}_c - \alpha {\bf K}_b $. Functions $f$ appearing
in the radial integrals ${\cal R}_{L_a,L_b,L_c}$ are the   
radial parts of the distorted wave functions $\chi$'s of Eq.~(2). These 
are calculated by solving the Schr\"odinger equation with appropriate optical
potentials, which include both the Coulomb and the nuclear terms.
The slowly converging integrals ${\cal R}_{L_b,L_c,L_a}$ can be handled
effectively by using the complex plane method~\cite{vin70}. 

This theory can be used to calculate breakup of both neutron
and proton halo nuclei. Generally, the maximum value of the partial
waves $L_a,L_b,L_c$ must be very large in order to ensure the convergence
of the partial wave summations in Eq.~(5). However, for the case of the
one-neutron halo nuclei, one can make use of the following method 
to include summations over infinite number of partial
waves. We write $\beta_{\ell m}$ as
\begin{eqnarray}
\beta_{\ell m} & = & \sum_{L_i = 0}^{L_{i}^{max}} {\hat \beta}_{\ell m} (L_i) 
             + \sum_{L_i = L_{i}^{max}}^{\infty} {\hat \beta}_{\ell m}(L_i),    
\end{eqnarray}
where ${\hat \beta}$ is defined in the same way as Eq.~(5) except for
the summation sign and $L_i$ corresponds to $L_a$, $L_b$, and $L_c$. If
the value of $L_i^{max}$ is chosen to be appropriately large, the
contribution of the nuclear fields to the second term of Eq.~(6) can be
neglected and we can write
\begin{eqnarray}
\sum_{L_i = L_{i}^{max}}^{\infty}{\hat \beta}_{\ell m}(L_i)  \approx    
            \sum_{L_i = 0}^{\infty}{\hat \beta}_{\ell m}^{Coul}(L_i) -   
             \sum_{L_i = 0}^{L_{i}^{max}}{\hat \beta}_{\ell m}^{Coul} (L_i),
\end{eqnarray}
where the first term on the right hand side, is the pure Coulomb 
breakup amplitude which for
the case where one of the outgoing fragments
is uncharged, can be expressed analytically in terms of the 
Bremsstrahlung integral (see, e.g.,~\cite{cha00}). Therefore, only
two terms, with reasonable upper limits, are required to be evaluated
by the partial wave expansion in Eq.~(6).

The wave function, $u_\ell(r)$, appearing in the structure term, $Z_\ell$,
has been calculated by adopting a single particle potential
model. The ground state of $^{11}$Be was assumed to have a 2$s_{1/2}$
valence neutron coupled to the $^{10}$Be$(0^+)$ core with a binding energy
of 504 keV and a spectroscopic factor of 0.78. The corresponding single
particle wave function was constructed by assuming the neutron - $^{10}$Be
interaction of a central Woods-Saxon type. For a given set of the radius and
diffuseness parameters (1.15 fm and 0.5 fm, respectively~\cite{cha00}),
the depth of this potential was searched so as to reproduce the ground
state binding energy. The neutron-target optical potentials used by us 
were extracted from the global set of Bechhetti-Greenlees
(see, e.g,~\cite{per76}), while those used for the $^{10}$Be + target
(~\cite{per76,beu98}) system are shown in Table I. Following~\cite{typ01},
we have used the sum of these two potentials for the
$^{11}$Be-target channel. We found that values of $L_i^{max}$ of 500
for Au and Ti and 150 for the Be provided very good convergence of the
corresponding partial wave expansion series [Eq.~(6)]. 

In Fig.\ 1, we show our results for the  neutron angular
distributions ($d\sigma/d\Omega_n$) for the reaction as mentioned in the
corresponding figure caption. Our calculations are in good
agreement with the experimental data~\cite{ann94} 
(shown by solid circles) for all the three targets. 
For the Be target, $d\sigma/d\Omega_n$
is governed solely by the nuclear breakup effects at all the angles.
The pure Coulomb breakup contributions are down by at least an
order of magnitude at the forward angles and by 2-3 orders of
magnitude at the backward angles. The CNI terms are also small
in this case.  On the other hand, for Ti and Au targets the Coulomb
terms are dominant at the forward angles while the nuclear breakup
effects are important at larger angles. Magnitudes of the CNI terms
vary with angle; for many 
\begin{table}[here]
\caption{Optical potential parameters for the $^{10}$Be-target
interaction. Radii are calculated with the $r_jt^{1/3}$ convention.}
\begin{tabular}{|c|c|c|c|c|c|c|}
\hline
system &  $V_r$ & $r_r$& $a_r$&  $W_i$& $r_i$& $a_i$\\
       & (Mev) & (fm) & (fm) & (Mev) & (fm) & (fm)       \\
\hline
$^{10}$Be--$^{197}$Au & 400 & 2.08 & 0.9 & 76.2 & 1.52 & 0.38  \\
$^{10}$Be--$^{44}$Ti &70 & 2.5 & 0.5 & 10.0 & 1.5 & 0.50 \\
$^{10}$Be--$^{9}$Be & 100 & 2.6 & 0.5& 18.0 & 2.6 & 0.50  \\
\hline
\end{tabular}
\end{table}
\noindent
forward angles they almost coincide with those
of the nuclear breakup while at the backward angles they are closer to the
pure Coulomb breakup contributions. Signs of these terms also change
with the neutron angle; a feature common to all the three targets.
It is clear
\begin{figure}[here]
\begin{center}
\mbox{\epsfig{file=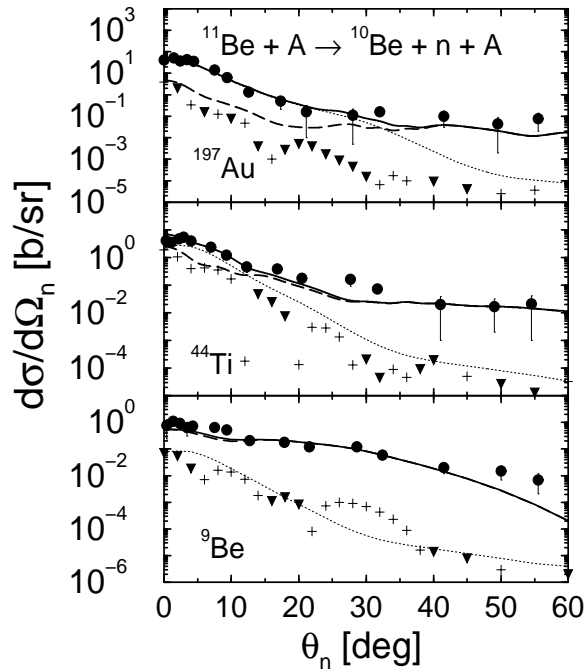,height=9.0cm,width=7.6cm}}
\end{center}
\vskip .1in
\caption {
Neutron angular distribution for the breakup reaction $^{11}$Be + A 
$\to$ $^{10}$Be + n + A at the beam energy of 41 MeV/nucleon. 
The dotted and dashed
lines represent the pure Coulomb and nuclear contributions, respectively
while their coherent sums are shown by the solid lines. The plus signs
and the inverted solid triangles represent the magnitudes of the 
positive and negative interference terms, respectively. 
The data are taken from~\protect\cite{ann94}.
}
\label{fig:figa}
\end{figure}
\noindent
that the interference terms are not negligible for
Ti and Au targets at the forward angles. For $\theta_n$
$\leq$ 10$^\circ$, the magnitudes of the CNI contributions are similar
to those of the pure nuclear terms, leading to a better description of the
data in this region.
 
In Fig.\ 2, we compare the results of our calculations with the 
data (taken from~\cite{nak94}) for the relative energy spectrum of 
the fragments (neutron and $^{10}$Be) emitted in the breakup
of $^{11}$Be on a  
$^{208}$Pb target at the beam energy of 72 MeV/nucleon. The optical
potential parameters, in this case, were taken to be the same as those
used for the gold target. We note that while the pure
Coulomb contributions dominate the cross sections around the peak value,
the nuclear breakup is important at the larger relative energies.
This is attributed to the different  energy dependence of the two
contributions~\cite{typ01}.
The coherent sum of the Coulomb and nuclear contributions provides a
good overall description of the experimental data. The nuclear and the
CNI terms are necessary to explain the data at larger relative energies.
Despite the peripheral nature of the reaction,
nuclear interactions between the projectile and the target may become
possible due to the extended nature of the $^{11}$Be wave function.
This is the reason for the failure of the pure Coulomb DWBA
calculations~\cite{cha00} in describing properly the cross sections 
in this region.

The effect of the interference terms is small (of the order of 2-8$\%$)
on the total breakup cross section.  It is constructive for the Be and
Ti targets and destructive for the Au target. Therefore, the role of
the CNI terms in the total breakup cross section is dependent on the
target nucleus.
\begin{figure}[here]
\begin{center}
\mbox{\epsfig{file=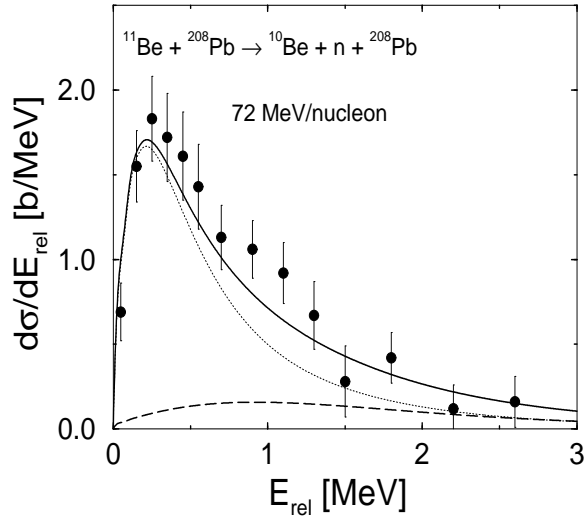,height=7.0cm,width=7.6cm}}
\end{center}
\caption {
The differential cross section as a function of the relative energy of the
fragments (neutron and $^{10}$Be) in the breakup reaction os $^{11}$Be
on a $^{208}$Pb target. Various curves have the same
meaning as that in Fig.\ 1. The data are taken from~\protect\cite{nak94}. 
}
\label{fig:figb}
\end{figure}

In summary, we have developed a complete quantal formulation for
investigating the breakup reactions of the halo nuclei within the
framework of the post form distorted wave Born approximation, where
the pure Coulomb, the pure nuclear as well as their interference
terms are treated consistently within the same framework. Our theory
takes into account both Coulomb and nuclear parts of the
fragment-target interactions to all orders, while the interaction
between the fragments is treated in first order. It may be mentioned 
that the higher order dynamical polarization processes that become
important at lower energies in the Coulomb dissociation of proton halo
nuclei~\cite{esb02} may not have been treated properly in our theory.
However,  this effect does not play any role for the Coulomb dissociation of
the neutron halo nuclei which is the subject of study in this paper. 
Nevertheless, the lack of proper knowledge of
the appropriate optical potentials, particularly in the halo
projectile-target channel is a source of uncertainty in our 
calculations which is indeed the case for all reaction studies 
of halo nuclei where the distorted waves in the projectile-target
channel are required.

As a first numerical application of this theory, we studied breakup
of the one-neutron halo nucleus $^{11}$Be on several target nuclei.
We calculated the angular distributions of the neutron fragment 
emitted in breakup reactions of this nucleus on Be, Ti and Au
target at the beam energy of 41 MeV/nucleon. The results of our
calculations are in good agreement with the available data for all
the three targets. 
We find that for medium mass and heavy target nuclei,
the neutron angular distributions 
are dominated by the nuclear and the Coulomb breakup
terms at larger and smaller angles, respectively. Contributions
of the Coulomb-nuclear interference terms are non-negligible.
They can be as big in magnitude as the pure nuclear or the pure
Coulomb breakup and have negative or positive sign depending upon
the angle and energy of the outgoing fragments. For these targets,
the interference terms help in better description of the trends of
the experimental data even at smaller angles. Similarly, the data on the
relative energy spectra of the fragments (neutron and $^{10}$Be)
emitted in breakup of $^{11}$Be on a Pb target at the beam energy of
72 MeV/nucleon, can not be described
properly by considering only the pure Coulomb breakup mechanism;
inclusion of the nuclear and Coulomb-nuclear interference terms is 
necessary. In most of the previous studies of this reaction, these
terms were neglected. Therefore, the exclusive halo breakup data on
medium mass and the heavy target nuclei need to be analyzed more
accurately than what has been done so far. 

More results on the comparison of calculations performed within this theory
and the halo breakup data, particularly on the momentum distribution
of fragments, will be presented elsewhere. 
Work is under way on the calculations of the breakup amplitude  
[Eq. (2)] without making the local momentum approximation (which is
computationally a very involved problem) so that 
the question of the validity of this approximation can be
addressed more rigorously.


\begin{references}
\bm{tan88}
I. Tanihata, J. Phys. G: Nucl. Part. Phys {\bf 22}, 157 (1996). 

\bm{han95}
P.G. Hansen, A.S. Jensen, and B. Jonson, Annu. Rev. Nucl. Part. Sci. {\bf 45},
2 (1995).

\bm{nak94}
T. Nakamura {\em et al.}, Phys. Lett. {\bf B331}, 296 (1994); 
{\em ibid}, Phys. Rev. Lett. {\bf 83}, 1112 (1999).

\bm{han87}
P.G. Hansen and B. Jonson, Europhys. Lett. {\bf 4}, 409 (1987). 

\bm{dav98}
B. Davids {\it et al.}, Phys. Rev. Lett. {\bf 81}, 2209 (1998); 
D. Cortina-Gil {\it et al.}, Eur. Phys. J. {\bf A10}, 49 (2001); 
R. Kanungo {\it et al.}, Phys. Rev. Lett. {\bf 88}, 142402 (2002).

\bm{ann94}
R. Anne {\em et al.}, Nucl. Phys. {\bf A575}, 125 (1994).;
R. Anne {\em et al.}, Phys. Lett. {\bf B250}, 19 (1990).

\bm{pra93}
P. Banerjee and R. Shyam, Nucl. Phys. {\bf A540}, 112 (1993),
{\em ibid}; J. Phys.G: Nucl. Part. Phys. {\bf 22}, L79 (1996).

\bm{mad01}
V. Maddalena {\em et al.}, Phys. Rev. C {\bf 63}, 024613 (2001),
and references therein.

\bm{mds01} 
V. Maddalena and R. Shyam, Phys. Rev. C {\bf 63}, 051601 (2001).

\bm{bau84}
G. Baur, F. R\"osel, D. Trautmann, and R. Shyam, Phys. Rep. {\bf 111}, 333
(1984).

\bm{sat64}
G.R. Satchler, Nucl. Phys. {\bf 55}, 1 (1964).

\bm{cha00}
R. Chatterjee, P. Banerjee, and R. Shyam, Nucl. Phys. {\bf A675}, 477 (2000).

\bm{cha01}
R. Chatterjee, P. Banerjee, and R. Shyam, Nucl. Phys. {\bf A692}, 476 (2001).

\bm{hen96}
G.F. Bertsch, K. Henken, and H. Esbensen, Phys. Rev. {\bf C57}, 1366 (1998);
H. Esbensen and G. Bertsch, Phys. Rev. {\bf C59}, 3240 (1999).

\bm{bon88}
A. Bonaccorso and F. Carstoiou, Phys. Rev. {\bf C61}, 034605 (2000) and
references therein.

\bm{yab92}
K. Yabana, Y. Ogawa, and Y. Suzuki, Nucl. Phys. {\bf A539}, 295 (1992).

\bm{shy99}
R. Shyam and I.J. Thompson, Phys. Rev. {\bf C59}, 2645 (1999).

\bm{kam86}
M. Kamimura, M. Yahiro, Y. Iseri, H. Kameyama, Y. Sakuragi, and M. Kawai,
Prog. Theor. Phys. Suppl. {\bf 89}, 1 (1986);
G.Baur, R. Shyam, F. R\"osel, and D. Trautmann, Phys. Rev. {\bf C28},
946 (1983).
 
\bm{tos01}
J.A. Tostevin, F.M. Nunes, and I.J. Thompson, Phys. Rev. {\bf C63}, 024617
(2001).

\bm{mor02}
J. Mortimer, I.J. Thompson, and J.A. Tostevin, Phys. Rev. {\bf C65},
064619 (2002); A.N. Moro, R. Crespo, F. Nunes, and I.J. Thompson,
Phys. Rev. {\bf C66}, 024612 (2002).

\bm{ian02}
I.J. Thompson, private communication.

\bm{typ01}
S. Typel and R. Shyam, Phys. Rev. {\bf C64}, 024605 (2001).

\bm{bon02}
J. Margueron, A. Bonaccorso, and D.M.  Brink, Nucl. Phys. {\bf A703}, 105
(2002)

\bm{bra74}
P. Braun-Munzinger and H.L. Harney, Nucl. Phys. {\bf A233}, 381 (1974);

\bm{pra02}
P. Banerjee, G. Baur, K. Hencken, R. Shyam, and D. Trautmann,
Phys. Rev. {\bf C65}, 064602 (2002).

\bm{gui00}
V. Guimar$\widetilde{a}$es {\it et al.}, Phys. Rev. {\bf C61}, 064609 (2000)

\bm{shy85}
R. Shyam and M.A. Nagarajan, Ann. Phys. (NY) {\bf 163}, 285 (1985).

\bm{zad02}
M. Zadro, Phys. Rev. C {\bf 66}, 034603 (2002).
 
\bm{vin70}
C.M. Vincent and H.T. Fortune, Phys. Rev. {\bf C2}, 782 (1970).

\bm{per76}
C. M. Perey and F. G. Perey, Atomic Data and Nuclear Data Tables {\bf 17},
1 (1976)

\bibitem{beu98}
M. Beunard {\it et al.}, Nucl. Phys. {\bf A424}, 313 (1984).

\bm{esb02}
H. Esbensen and G.F. Bertsch, Nucl. Phys. {\bf A706}, 383 (2002);
H. Esbensen and G.F. Bertsch, Phys. Rev. {\bf C66}, 044609 (2002).
\end{references}
\end{document}